\documentclass{eas}
\usepackage{graphicx}
\usepackage{natbib}
\newcommand{\eg} {{\em e.g.}}
\newcommand{\emm}[1]{\ensuremath{#1}}   % ensures math mode.
\newcommand{\emr}[1]{\emm{\mathrm{#1}}} % uses math roman fonts.
\newcommand{\thCO}{\emr{^{13}CO}}
\newcommand{\twCO}{\emr{^{12}CO}}
\newcommand{\HCOp}{\emr{HCO^+}} 
\newcommand{\HI}{\emr{HI}} 
\newcommand{\HH}{\emr{H_2}}
\newcommand{\Cp}{\emr{C^+}}
\newcommand{\Jone}{\mbox{(J=1--0)}}
\newcommand{\N}[1]{\emr{N_{#1}}}
\newcommand{\NH}{\N{H}}
\newcommand{\NHI}{\N{HI}}
\newcommand{\NHH}{\N{H_2}}

\newcommand{\NCO}{\N{CO}}

\newcommand{\W}[1]{\emm{W_\emr{#1}}}
\newcommand{\WCO}{\W{CO}}
\newcommand{\WCOperp}{\W{CO_\perp}}
\newcommand{\X}[1]{\emm{X_\emr{#1}}}
\newcommand{\XCO}{\X{CO}}
\newcommand{\f}[1]{\emm{f_\emr{#1}}}
\newcommand{\fHI}{\f{HI}}
\newcommand{\fHH}{\f{\emr{H_2}}}
\newcommand{\mytau}[1]{\emm{\int \tau_\emr{#1}\,dv}}
\newcommand{\tauhi}{\mytau{HI}}

\newcommand{\zoph}{\emm{\zeta{}}Oph}
\newcommand{\mean}[1]{\emm{ \left<  #1 \right> }}
\newcommand{\Rsun}{\emm{R_\odot}}
\newcommand{\Ebv}{\emm{E_\emr{B-V}}}
\newcommand{\Av}{\emm{A_\emr{V}}}
\newcommand{\unit}[1]{\emm{\, \emr{#1}}}
\newcommand{\pc}{\unit{pc}}
\newcommand{\kpc}{\unit{kpc}}
\newcommand{\K}{\unit{K}}
\newcommand{\kms}{\unit{km\,s^{-1}}}
\newcommand{\Kkms}{\unit{K\,km\,s^{-1}}}
\newcommand{\pccm}{\unit{cm^{-3}}}
\newcommand{\pscm}{\unit{cm^{-2}}}
\newcommand{\magn}{\unit{mag}}
\newcommand{\about}{\emm{\sim}}
\newcommand{\abs}[1]{\emm{\left| #1 \right|}}
\newcommand{\paren}[1]{\emm{\left(  #1 \right)}}
\renewcommand{\mean}[1]{\emm{ \left<  #1 \right> }}
\newcommand{\ga}{\mbox{\rlap{\hbox{\lower4pt\hbox{$\sim$}}}\hbox{$>$}}} % About lower than.
\newcommand{\la}{\mbox{\rlap{\hbox{\lower4pt\hbox{$\sim$}}}\hbox{$<$}}} % About lower than.
\begin{document}

%%-----------------------------
%%      the top matter
%%-----------------------------
\title{The CO-\HH\ conversion factor of diffuse ISM: \\
  Bright $^{12}$CO emission also traces diffuse gas} 
\author{J. Pety}\address{Institut de Radioastronomie Millim\'etrique \&
           Obs. de Paris, France, \texttt{pety@iram.fr}}
\author{H. S. Liszt}\address{National Radio Astronomy Observatory, USA, \texttt{hliszt@nrao.edu}}
\author{R. Lucas}\address{Joint ALMA Observatory, Chili, \texttt{rlucas@alma.cl}}
\thanks{The authors acknowledge funding by the grant ANR-09-BLAN-0231-01
  from the French {\it Agence Nationale de la Recherche} as part of the
  SCHISM project.}

\begin{abstract}
  We show that the \XCO{} factor, which converts the CO luminosity into the
  column density of molecular hydrogen has similar values for dense, fully
  molecular gas and for diffuse, partially molecular gas. We discuss the
  reasons of this coincidence and the consequences for the understanding of
  the interstellar medium.
\end{abstract}

\maketitle
%%-----------------------------
%%      your text
%%-----------------------------

\vspace*{-0.25cm}
\section{Introduction: Dense or diffuse gas?}
\vspace*{-0.25cm}

\twCO{} \Jone{} emission is the main tracer of the molecular gas in our
Galaxy as well as in external galaxies~\citep[see
\eg{}][]{LerWal+08,BigLer+08}. Yet, this CO emission is ambiguous. Indeed,
this CO emission is usually associated to cold (10-20\K), dense ($>
10^5\pccm$), strongly UV-shielded, molecular gas (\ie{} carbon atoms are
mostly locked in CO). However, it became more and more obvious over the
last decade~\citep{LisLuc98,GolHey+08} that a large fraction of the CO
emission in our Galaxy comes from warm (50-100\K{}), low density
(100-500\pccm), weakly UV-shielded, diffuse gas (\ie{} carbon atoms are
mostly locked in \Cp{}).  Moreover, at the beginning of the history of the
\XCO{} factor, \citet{Lis82,YouSco82} noted that diffuse and dense gas have
similar factor. For instance, $\WCO \approx 1.5\Kkms$, $\NHH{} = 5\times
10^{20}\,\kms$ toward \zoph{}, a prototypical diffuse line of sight, and
$\WCO = 450\Kkms$, $\NHH{} = 2\times 10^{23}\HH$ toward Ori A, a
prototypical dense gas line of sight. This triggers the question of the
mean value of the \XCO{} factor in diffuse gas~\citep[for details see]{LisPet10}.

\begin{figure}
  \centering
  \includegraphics[width=\textwidth]{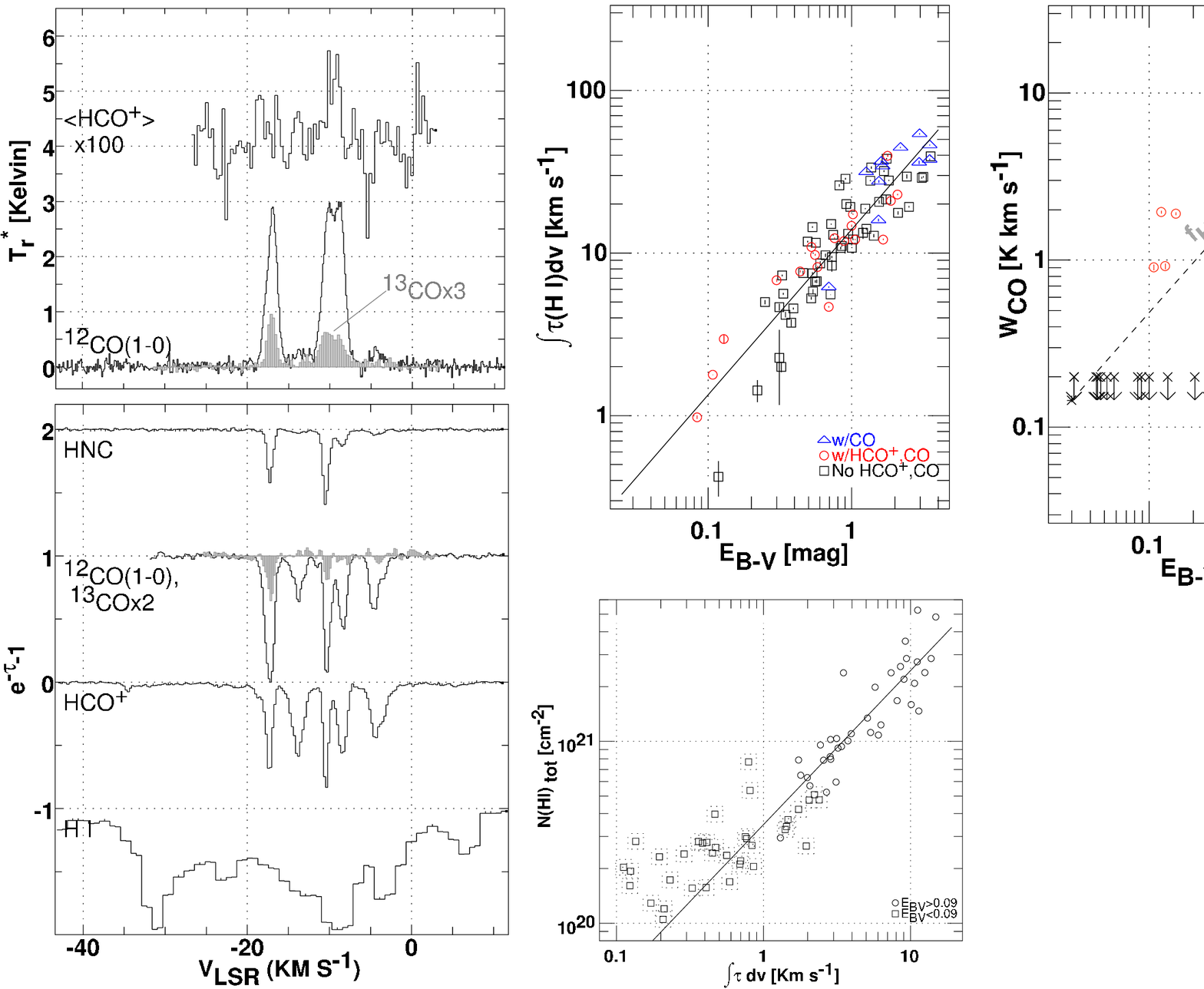}
  \caption{\textbf{Left:} Example of the measurements available in emission (top)
    and in absorption (bottom) for one of the Galactic line of sights used
    in this study~\citep[for details see][]{PetLuc+08}. \textbf{Middle,
      top:} Integrated VLA \HI{} optical depth from \citet{GarDic89} and
    this work versus the total line of sight reddening from
    \citet{SchFin+98}.  \textbf{Middle, bottom:} Total hydrogen column
    density versus integrated \HI{} optical depth for the sources studied
    by \cite{HeiTro03}. \textbf{Right:} CO luminosity versus the total line
    of sight reddening from \citet{SchFin+98}.}
  \label{fig:pety:1}
\end{figure}

\vspace*{-0.25cm}
\section{How to measure the mean $\NHH/\WCO$ conversion factor in 
  diffuse gas?}
\vspace*{-0.25cm}

We quantify the \XCO{} conversion factor in diffuse gas from a sample
acquired over the last 20 years. It is made from the study of whole
Galactic lines of sight measured in absorption against extragalactic
continuum background sources. There are two different groups of lines:
Either they have a low visual extinction $(\Av \la 1)$ when they are
observed at high galactic latitude $(\abs{b} \ga\,\, 15-20\deg)$ or when
they lie in the Galactic plane, their large total visual extinction $(\Av
\about 5)$ can be divided in well separated velocity components, each one
having $\Av \la\,\, 1\magn$ (see the left column of Fig.~\ref{fig:pety:1}
for an example).  Hence, all the gas studied here is diffuse.  Indeed, the
CO column density per velocity component on any line of sight is low,
typically $\NCO \le 2 \times 10^{16}\pccm$, which implies that less than
7\% of the carbon is locked in CO.

The computation is then made in three steps. First, the total hydrogen
column density is deduced from the \Ebv{} reddening maps
of~\citet{SchFin+98}, using the standard
relation~\citep{BohSav+78,RacSno+09}
\begin{equation}
  \NH = \NHI+2\NHH = 5.8\times 10^{21}{\rm H}\pscm\Ebv.
\end{equation}
Second, we estimate the atomic gas fraction via the \HI{} absorption
measurements by writing it as the product of two terms
\begin{equation}
  \mean{\fHI} = \mean{\frac{\NHI}{\NH}} \sim
        \mean{\frac{\NHI}{\tauhi}} \times \mean{\frac{\tauhi}{\NH}},
\end{equation}
where the left term is directly measured from the data (middle, top panel
of Fig.~\ref{fig:pety:1}) while the right term is calibrated from the
careful measurements of \citet{HeiTro03} (middle, bottom panel of
Fig.~\ref{fig:pety:1}). This gives $\mean{\fHI} = 0.65$ or $\mean{\fHH} =
2\NHH/\NH = 0.35$\footnote{Others measured in diffuse gas gives $0.25 \le
  \fHH \le 0.45$, which implies a 30\% overall uncertainty on the method.}.
The third step consists in the measures of the CO emission luminosity,
\WCO{}, along the line of sight (right panel of Fig.~\ref{fig:pety:1}). We
note here that the high \WCO{} values $(> 10\K)$ arise from the
accumulation of several low-\Av{} components along the same Galactic line
of sight. Moreover, CO is not reliably detected at low \Ebv{} $(<
0.3\magn)$. The corresponding lines of sight are not used in our estimation
because the \XCO{} factor is used to estimate the molecular column density
only from the CO gas which is detected. In summary, we obtain $\mean{\Ebv}
= 0.89\magn$ , $\mean{\fHH}=0.35$ and $\mean{\WCO} = 4.4\Kkms$, which give
$\NHH/\WCO=2.04 \times 10^{20}\HH\pscm/(\Kkms)$, \ie{} the same \emph{mean}
CO luminosity per \HH{} in diffuse and dense gas!

\vspace*{-0.25cm}
\section{Why a common $\NHH/\WCO$ conversion factor for diffuse 
  and dense gas?}
\vspace*{-0.25cm}

To understand this result, we write
\begin{equation}
  \frac{\NHH}{\WCO} = \paren{\frac{\NHH}{\NCO}}\,\paren{\frac{\NCO}{\WCO}},
\end{equation}
where the $\NHH/\NCO$ ratio comes from the CO chemistry while the
$\NCO/\WCO$ comes from the radiative transfer through the cloud structure.

In diffuse gas, more than 90\% of the carbon is locked in \Cp{}, which
implies that $\mean{\NCO/\NHH} = 3\times10^{-6}$~\citep{BurFra+07}. In
addition, the gas is subthermally excited. Large velocity gradient
radiative transfer methods~\citep{GolKwa74} thus show that 1) $\WCO/\NCO$
is large because of weak CO excitation in warm gas (60-100\K{}), and 2)
$\WCO \propto \NCO$ until the opacity is so large that the transition
approaches thermalization. We thus obtain
$\NCO/\WCO\simeq10^{15}\mbox{CO}\pscm/(\Kkms)$~\citep[see also][]{Lis07}.

In dense gas, all the carbon is locked in CO, \ie{} $\mean{\NCO/\NHH} =
10^{-4}$. As a consequence, $\NHH/\WCO$ = cst and $\WCO \propto \NCO$. A
constant \XCO{} factor thus implies that $\WCO \propto \NCO$. We interpret
this as a bulk effect in a turbulent medium, \ie{} the medium is
macroscopically optically thin because of the large velocity gradients due
to turbulence.

In summary, the change of chemistry from diffuse to dense gas is
compensated by the inverse change of the radiative transfer, giving the
same \XCO{} factor.

\vspace*{-0.25cm}
\section{How to discriminate diffuse from dense gas?}
\vspace*{-0.25cm}

As \twCO{} alone can not be used to discriminate between diffuse and dense
gas, other tracers as molecules with higher dipole moments (\eg{} \HCOp{},
CS, HCN) are tried. However, they often are difficult to detect. We thus
propose to use the CO isotopologues, \ie{} the ratio of the \twCO{} over
the \thCO{} emission. Indeed, $T_{\twCO}/T_{\thCO} \ga\,\, 10-15$ in
diffuse, warm gas because of the \Cp{}
fractionation~\citep[\eg{}][]{LisLuc98}, while $T_{\twCO}/T_{\thCO} \la\,\,
3-5$ in dense, cold gas~\citep[\eg{}][]{BurGor78}.

\vspace*{-0.25cm}
\section{What is the proportion of CO emission arising from diffuse gas in
  our Galaxy?}
\vspace*{-0.25cm}

We here aim at estimating the CO luminosity of the diffuse molecular gas
perpendicular to the Galactic plane from our absorption data for which the
mean luminosity is $\mean{\WCO} = 4.6\Kkms$. We assume that the gas is
simply ordered in plane-parallel, stratified layers. The Mean number of
galactic half-width along integration path is then $\mean{1/\sin\abs{b}} =
19.8$, where $b$ is the latitude of each line of sight of our sample. We
thus obtain $\mean{\WCOperp} = 2\mean{\WCO(b)}/\mean{1/\sin\abs{b}} =
0.47\Kkms$.

The CO emission Galactic survey of~\citet{BurGor78} gives a mean CO
brightness per kpc of $5\Kkms/\kpc$ at $\Rsun = 8\kpc$. Assuming a single
Gaussian vertical distribution of dispersion 60\pc{}, we obtain
$\mean{\WCOperp} = 0.75\Kkms$. This is a lower limit because we see large
amount of CO emission outside area mapped by the Galactic surveys of the CO
emission (mainly limited to the Galactic plane). Hence, a large fraction of
the CO luminosity measured in Galactic surveys could come from diffuse gas.

\vspace*{-0.25cm}
\section{Conclusion: Interpreting a sky occupied by CO emission from
  diffuse gas}
\vspace*{-0.25cm}

The coincidence between the values of the \XCO{} factor in diffuse and
dense gas implies that the mass estimates computed with the standard value
are correct. However, the underlying physical interpretation of the
detected gas is very different. If the gas is dense: it will fill a small
fraction of the interstellar volume; it will be confined by ram or
turbulent pressure (if not gravitationally bound); and it is on the verge
of forming stars. If the gas is diffuse: it is a warmer, low pressure medium
filling a large fraction of the interstellar volume; it contributes more
the mid-IR or PAH emission; and it is probably not gravitionally bound or
about to form stars.

%%-----------------------------
%%      your bibliography
%%-----------------------------

%%\bibliographystyle{apj} \bibliography{mnemonic,absorption}

\end{document}